\documentclass[showpacs,twocolumn,superscriptaddress,nofootinbib]{revtex4}

\usepackage{doi}
\usepackage{hyperref}
\hypersetup{
  colorlinks=true,        %
  linkcolor=blue,         %
  citecolor=cyan,         %
}

\usepackage{graphicx}
\usepackage{dcolumn}
\usepackage{bm}
\usepackage{color}
\usepackage{footmisc}

\usepackage{amsmath}
\usepackage{amssymb}


\begin{document}

\title{The higher dimensional Myers-Perry black hole with single rotation always obeys the Cosmic
Censorship Conjecture} 

\author{Sanjar Shaymatov}
\email{sanjar@astrin.uz}

\affiliation{Ulugh Beg Astronomical Institute, Astronomicheskaya
33, Tashkent 100052, Uzbekistan}

\affiliation{Institute of Nuclear Physics, Ulughbek, Tashkent 100214, Uzbekistan}

\author{Naresh Dadhich}
\email{nkd@iucaa.in}

\affiliation{Inter University Centre for Astronomy \&
Astrophysics, Post Bag 4, Pune 411007, India }

\author{Bobomurat Ahmedov}
\email{ahmedov@astrin.uz}

\affiliation{Ulugh Beg Astronomical Institute, Astronomicheskaya
33, Tashkent 100052, Uzbekistan}
\affiliation{ National University
of Uzbekistan, Tashkent 100174, Uzbekistan }

\date{\today}
\begin{abstract}
{Even though the Myers-Perry five dimensional rotating black hole with two rotations could be overspun
by test particle accretion, yet it turns out as we show in this letter that it cannot do so for a single rotation. 
On the other hand it is known that there exists no extremal limit for a black hole with single rotation in dimensions 
greater than equal to six. It has been proven that all 
higher dimensional ($>4$) rotating black holes with only one single rotation can never be overspun 
under test particle linear accretion and hence would always obey CCC in the weak form. }

\end{abstract}
\pacs{04.50.+h, 04.20.Dw} \maketitle

\section{Introduction}
\label{introduction}

The occurrence of singularity is a generic property of Einstein's
gravity -- general relativity (GR), which has been established by
very powerful singularity theorems \cite{Hawking-Penrose70}. A
singularity marks the limit of a theory's applicability where it
loses its predictive power. Fortunately, singularity that results
under gravitational collapse of an object that has exhausted all
its resources to counter gravity, is covered by an event horizon
of black hole. Hence it turns benign for the region lying outside
black hole horizon because no signal could come out of horizon.

For validity and applicability of the Einstein gravity in the
large, Penrose pronounced in 1969 \cite{Penrose69} that this would
always be the case in general -- a singularity would always be
hidden behind a horizon. This is what is called the Cosmic Censorship
Conjecture (CCC) (all through by CCC we would mean weak CCC).
There exists no proof of the conjecture either
way, true or false, and it remains as one of the most important
open questions. It has two aspects, one whether gravitational
collapse can never end in a naked singularity without a horizon
cover, and two could an existing horizon of a charged/rotating
black hole be destroyed by over-charging/rotating? Even though
there exists a vast literature on the former
\cite{Joshi93,Joshi00,Joshi15,Stuchlik12a,Vieira14,Stuchlik14},
yet there is no consensus on the occurrence of naked singularity.
However for generic conditions collapse of a differentially
rotating neutron star suggests that CCC is not
violated~\cite{Giacomazzo-Rezzolla11}. We shall however focus here
on the latter aspect of destroying horizon of a near extremal
black hole by impinging it with test particles of suitable
parameter in a Gedanken experiment.

First the question addressed was, whether a non-extremal black
hole could be turned extremal by particle accretion leading to
black hole temperature reducing to zero -- violation of the third
law of black hole thermodynamics? It was shown
\cite{Wald74b,Dadhich97} that it was not possible because as
extremality was approached the particle parameter space pinched
off -- particles with suitable parameters won't be able to reach
horizon near extrtemality. Then the question was formulated
somewhat differently, true, extremality may not be achievable,
could it however be jumped over? That is, in a discontinuous
manner a black hole could jump from sub-extremal to over-extremal
state, without passing through extremality -- rather jumping over
it, thereby destroy horizon and create a naked singularity.

A near extremal charged black hole $Q<M$ was shown \cite{Hubeny99}
to be overcharged to $Q>M$ state by accretion of overcharged test
particles. It was a linear order process in which effects of self
force as well as of finite size of particle were ignored.
Following that the same result was extended to rotating black hole
\cite{Jacobson09,Saa11} --  a rotating black hole could similarly be
overspun. Following \cite{Hubeny99,Jacobson09} there have been several
investigations on these lines \cite{Bouhmadi-Lopez10,Li13,Rocha14,
Shaymatov15,Song18,Duztas18} addressing the question of 
over-charging/spinning of black hole.\footnote{In Ref. \cite{Bouhmadi-Lopez10}, it 
is essentially shown for different five dimensional black hole geometries 
that geodesic particle accretion can never lead to extremality. This is because 
parameter space required for attaining extremality pinches off as it is approached 
\cite{Wald74b,Dadhich97}. { Very recently this calculation is also done for multi 
black hole system \cite{Mishra19} showing the same result.}} 

It turns out that horizon of a charged and or rotating black hole can be
destroyed by particle accretion of suitable parameters, and a naked singularity
can be created. In all these works, it was assumed that
test particle follows a geodesic {(Lorentz force trajectory for charged particle)} motion, and contributions of backreaction,
self force and  radiative effects were not taken into consideration.
If these effects are taken into account,  particles that could overspin the
black hole would not be able to fall into the horizon
~\cite{Barausse10,Rocha11,Isoyama11,Zimmerman13,Colleoni15a,Colleoni15b}. And so
the CCC may indeed be respected.

All this was in the linear regime, and the question was open
whether the result would stand when non-linear accretion regime was
considered. That has recently been done \cite{Sorce-Wald17} and
the answer has come out in favor of CCC, that black hole cannot be
over-charged/spun -- horizon cannot be destroyed. It has also been
shown that the same is the case for Kerr-AdS black hole \cite{Gwak18a}.

Another question then arises, what happens in higher dimensions --
could a five dimensional black hole be overcharged or overspun? It
is shown \cite{Revelar-Vega17} that a five dimensional charged
black hole could be overcharged for linear test particle accretion and the horizon is destroyed.
Following \cite{Sorce-Wald17}, there have been some non-linear accretion
studies \cite{Ge18} verifying the validity of CCC. {The general picture 
that emerges is that for linear order accretion a black hole could be overcharged or 
overspun and CCC is violated while it could not be done so for non-linear accretion and 
thereby restoring CCC.}  

A rotating black hole in higher dimensions has more than one rotation
axis; i.e. a five dimensional black hole could have two rotation
parameters about the two axes. {Linear as well as non-linear accretion
process have been studied \cite{An18} for the five dimensional
Myers-Perry rotating black hole \cite{Myers-Perry86}, and it is shown
that though CCC is violated for the linear accretion but it is as expected obeyed
for the non-linear accretion. This was all done for black hole having two rotation parameters.  
Could the situation be different if black hole has only one single rotation? The answer to this 
question is surprisingly yes for a five dimensional rotating black hole under linear accretion. 
To investigate this question defines the purpose of this letter.}

{In this letter we wish to point out that the situation is radically
different for a five dimensional black hole having a single rotation than
the one having two rotations under linear accretion. It remarkably turns out that a black hole
with single rotation could not be overspun for linear test particle
accretion, and thereby CCC is respected. This is in stark contrast with
the four dimensional Kerr as well as the five dimensional Myers-Perry
black hole with two rotations \cite{An18} where it could be overspun and CCC is violated
for linear accretion. Besides our independent calculations, we have also verified
the result by switching off one of rotation parameters in the Ref. \cite{An18}
analysis which considers both linear and non-linear accretion for two rotations.} 
{The main result that emerges is that a five dimensional black hole with
only one rotation cannot be overspun even at linear accretion. As expected, this result continues 
to be true for non-linear accretion. This is different from 
all other cases where at linear order overspinning is always possible. That is, CCC is violated at 
linear order which is restored only when non-linear accretion is invoked.}  

{With this background the result stands out that five dimensional black hole with single rotation does 
not require non-linear accretion for respecting CCC. Hence CCC is always respected for single rotation five dimensional 
black hole irrespective of accretion being linear or non-linear. }

This happens because the minimum threshold angular momentum required
for overspinning turns out to be greater than maximum threshold allowed
for particle reaching horizon. Thus there is no parameter space available
for test particles that could lead to overspinning of black hole. This is exactly what
happens for extremal black hole, minimum threshold leading to over-extremality
turns out to be greater than the corresponding maximum threshold. That is why
an extremal horizon can never be destroyed -- an extremal black hole can never be
pushed to over-extremal state of naked singularity.

{The paper is organized as follows: In Sec.~\ref{Sec:metric}, we
briefly recall the five dimensional rotating black hole metric
which is followed by discussion of over extremality of black hole and we show that five dimensional rotating black hole with a single rotation cannot be
overspun even in linear accretion process in the
Sec.~\ref{Sec:overspinning}. We end up with conclusion in the
Sec.~\ref{Sec:conclusion}.}

\section{Five dimensional Myers-Perry rotating black hole}\label{Sec:metric}

{Further it is known \cite{Myers-Perry86} that a rotating black hole with a single
rotation has no extremal limit in dimension $\geq6$, and hence it cannot be overspun
simply because it can have arbitrary angular momentum without destroying horizon. Combining this fact 
with the above result we arrive at a very important and interesting conclusion that a rotating black hole with
single rotation in dimension greater than four always obeys CCC. We begin as follows:}

Let's recall the metric of the five dimensional Myers-Perry rotating black hole
without cosmological parameter $\Lambda$ \cite{Hawking99},

\begin{eqnarray}\label{5D_metric}
ds^2&=&-\frac{\Delta}{\Sigma}\left(dt-a\sin^2\theta
d\phi-b\cos^2\theta
d\psi\right)^2+\frac{\Sigma}{\Delta}dr^{2}\nonumber\\
&+&\Sigma d\theta^2 +
\frac{\sin^2\theta}{\Sigma}\left[(r^2+a^2)d\phi-a
dt\right]^2\nonumber\\&+&\frac{\cos^2\theta}{\Sigma}\left[(r^2+b^2)d\psi-b
dt\right]^2 \nonumber\\&+&
r^2\left(\cos^2\theta+\sin^2\phi\right)d\psi^2\ ,\ \quad
\end{eqnarray}
where $\Delta=\frac{(r^{2}+a^{2})(r^{2}+b^{2})}{r^{2}}-\mu$ and
$\Sigma=r^{2}+a^{2}\cos^{2}\theta+b^{2}\sin^{2}\theta$. Here $a=\frac{4J_{\phi}}{\pi\mu}$
and $b=\frac{4J_{\psi}}{\pi\mu}$ are rotation parameters about the two rotation axes, and
{${\mu=\frac{8M}{3\pi}}$} is mass parameter.

The black hole horizon is given by
 \begin{eqnarray}
 r_{\pm}&=&\left(\frac{1}{2}\right)^{1/2}\left[\left({\mu}-a^{2}-b^{2}\right)\right.\nonumber\\
 &\pm & \left.\sqrt{\left({\mu}-a^{2}-b^{2}\right)^{2}-4a^2b^{2}} \right]^{1/2},
 \end{eqnarray}

and extremality is indicated by $a + b = (\mu)^{1/2}$.

The necessary and sufficient conditions for
over-extremality(spinning) are: (a) An overspinning particle must
fall into black hole; i.e. it reaches horizon. {This will define
the maximum threshold for particle angular momentum $\delta
J_{max}$ as given in Eqs. (\ref{Jmax}-\ref{Jmax1}). This is the maximum angular momentum a particle 
can carry while falling into black hole. (b) On accretion resulting state of black hole should be
over-extremal/spun, that would define the minimum threshold,
$\delta J_{min}$} as given in Eqs. (\ref{naked}-\ref{naked1}). The parameter window $ {\Delta J} = \delta
J_{max} - \delta J_{min}$ defines the parameter space of accreting
particle required for overspinning. A black hole can overspin if
and only if ${\Delta J} > 0$, and if the opposite, ${\Delta J}<0$,
is true, it cannot be. In that case overspinning particle
would not be able to reach horizon to fall into black hole. This
is precisely what we shall show in the following that $\Delta J <0$ for a black hole having only one rotation parameter, and hence
black hole cannot be overspun.

\section{Overspinning of five dimensional rotating black hole with test particles}\label{Sec:overspinning}

Let a particle of energy ${\delta E }$ and angular momenta $\delta J= \delta J_{\phi}+ \delta
J_{\psi}$,
fall into a black hole. For particle to reach horizon, we have $\delta E \geq \Omega_{+}^{(\phi)}\delta
J_{\phi}+\Omega_{+}^{(\psi)}\delta J_{\psi}$, and so we write
\begin{equation}\label{Jmax}
\delta E \geq \frac{a}{r^2_{+}+a^2}~\delta J_{\phi}+\frac{b}{r^2_{+}+b^2}~\delta J_{\psi}\, ,
\end{equation}
where $\Omega_{+}^{(\phi)}$ and $\Omega_{+}^{(\psi)}$ are respectively two black hole angular
velocities relative to $\phi$ and $\psi$ axes.

By writing $\delta J_{\phi}=\lambda~\delta J $, $\delta J_{\psi}=\alpha~\delta J $ with
$\alpha+\lambda=1$,   the maximum threshold is defined by
\begin{eqnarray}\label{Jmax1}
 \delta
J_{max}=\frac{\mu r^2_{+}}{a\left(r^2_{+}+b^2\right)(1-\alpha)+b\left(r^2_{+}+a^2\right)\alpha} \delta
E\, .
\end{eqnarray}
On the other hand minimum threshold would be given by
\begin{equation}\label{naked} \sqrt{\frac{8}{3\pi}}\left({M+\delta E}\right)^{1/2}<
\frac{3}{2}\left(\frac{J_{\phi}+\delta J_{\phi}}{{M}+\delta E}+\frac{J_{\psi}+\delta
J_{\psi}}{{M}+\delta E}\right)\, ,
\end{equation}
\\
and hence
\begin{eqnarray}\label{naked1}
\delta
J_{min}&=& \delta J_{\phi}+\delta
J_{\psi}=\left(\sqrt{\frac{32}{27\pi}}{M}^{3/2}-J_{\phi}-J_{\psi}\right)\nonumber\\&+&\sqrt{\frac{32}{27\pi}}\left(\frac{3}{2}
M^{1/2}
\delta E
+\frac{3}{8}M^{-1/2}\delta E^2\right)\, .
\end{eqnarray}
Note that an exactly extremal black hole can never be
over-extremalized simply because in that case
$\delta J_{max} < \delta J_{min}$ leaving no parameter window
available for over-extremality. This is why
one has always to begin with a near extremal state, {$J_{\phi} +
J_{\psi} =\frac{2}{3}\left(a+b\right)M= \sqrt{\frac{32}{27\pi}}M^{3/2} \left(1-
\epsilon^2\right)$, $\epsilon\ll 1$.}
We write $a=\sqrt{\frac{8M}{3\pi}}\gamma \left(1-\epsilon^2\right)$ and
$b=\sqrt{\frac{8M}{3\pi}}\beta \left(1-\epsilon^2\right)$ with $\gamma+\beta=1$, then
$\delta J_{max}$ and $\delta J_{min}$ are written as

\begin{eqnarray}\label{jmax}
\delta J_{max}&=&
\left[1+\sqrt{2\frac{\left(\alpha-\beta\right)^2}{\left(1-\beta\right)\beta}}~\epsilon\right.\nonumber\\
&+&\left.\frac{\left[2\alpha^2\left(1-2\beta\right)^2-\alpha\left(1-6\beta+8\beta^2\right)+\beta^2\right]
}{\left(1-\beta\right)\beta}~\epsilon^2\right]\nonumber\\&\times& \sqrt{\frac{8}{3\pi}}~M^{1/2}~\delta
E\, ,
\end{eqnarray}

\begin{eqnarray}\label{jmin}
\delta J_{min} &=&\delta J_{\phi}+\delta J_{\psi}=\sqrt{\frac{8}{3\pi}}\left(
{\frac{2}{3}}~M^{3/2}~\epsilon^2\right.\nonumber\\&&+\left. M^{1/2}~\delta E +
{\frac{1}{4}}~M^{-1/2}~\delta E^2\right)\, ,
\end{eqnarray}

and

{\begin{eqnarray} \label{jmax-jmin}
\Delta J &=& \sqrt{\frac{8}{3\pi}}~\left(\left[\sqrt{2\frac{\left(\alpha-\beta\right)^2}{\left(1-\beta\right)\beta}}~\epsilon
\right.\right.\nonumber\\&+&\left.\frac{\left[2\alpha^2\left(1-2\beta\right)^2-\alpha\left(1-6\beta+8\beta^2\right)+\beta^2\right]
}{\left(1-\beta\right)\beta}~\epsilon^2 \right]\nonumber\\&\times& \left. M^{1/2}\delta
E -{\frac{2}{3}}M^{3/2}\epsilon^2- {\frac{1}{4}}M^{-1/2}\delta E^2 \right)\, .
\end{eqnarray}}

For overspinning, $\Delta J >0$ which would define parameter window for $\delta E$ as given by

\begin{eqnarray}\label{range}
&&2\left(\sqrt{2\frac{\left(\alpha-\beta\right)^2}{\left(1-\beta\right)\beta}}-\sqrt{2\frac{\left(\alpha-\beta\right)^2}{\left(1-\beta\right)\beta}-\frac{2}{3}}\right)
< \Delta E\nonumber\\&&<
2\left(\sqrt{2\frac{\left(\alpha-\beta\right)^2}{\left(1-\beta\right)\beta}}+\sqrt{2\frac{\left(\alpha-\beta\right)^2}{\left(1-\beta\right)\beta}-\frac{2}{3}}\right)
 \, ,
\end{eqnarray}
where $\Delta E={\delta E}/{M\epsilon}$. This shows that $\Delta J >0$ and there exists a parameter
window available as shown above for overspinning the black hole. A black hole with two rotations could
indeed be overspun. Thus the horizon would be destroyed leading to naked singularity, and CCC would be
violated. This
was what shown in Ref. \cite{An18}.

Let's first consider the case of infalling particle having only one angular momentum, $J_{\psi}$; i.e
$\alpha=1$, and black hole having the two rotations, then the above inequality would require $\beta <
3/4$. Again there would be parameter window available for test particles for overspinning the black
hole. That is, a black hole with two rotations could always be overspun so long as $\beta < 3/4$.
Clearly there would be no overspinning possible for $\beta > 3/4$. Note that single rotation parameter
means $\beta = 1$ indicating that black hole cannot be overspun in that case. However, we cannot take
this limit in the above inequality because the terms diverge, and hence we have to consider the case of
single rotation afresh separately. That is what we now do.

Let's begin by defining $\delta J_{min}$ for a single rotation, one can write
$\sqrt{\frac{32}{27\pi}}\left(M + \delta E\right)^{3/2} < J
+ \delta J$ and similarly for $\delta J_{max}$.
So the minimum threshold would be defined by
\begin{eqnarray}\label{Jmina=0}
\delta J_{min} &=& \sqrt{\frac{8}{3\pi}}\left({\frac{2}{3}}~M^{3/2}~\epsilon^2 + M^{1/2}~\delta E \right.
\nonumber\\
&&+ \left.{\frac{1}{4}}~M^{-1/2}~\delta E^2\right)\, .
\end{eqnarray}

Since $r_{+}=(\mu-b^2)^{1/2}$ and $b=\sqrt{\frac{8M}{3\pi}}(1-\epsilon^2)$ for a single rotation,
$\delta J_{max}$ yields
\begin{equation}\label{Jmaxa=0}
\delta
J_{max}=\frac{r^2_{+}+b^2}{b} \delta E=\frac{\mu}{b}~\delta E\, ,
\end{equation}
which, in turn, gives the upper bound up to $O(\epsilon^2)$,
\begin{equation}\label{Jmaxa=0} \delta
J_{max}=\sqrt{\frac{8}{3\pi}}\left(1+\epsilon^2\right)~M^{1/2} \delta E.
\end{equation}
So
\begin{eqnarray}\label{minmax}
 \Delta J &=&\sqrt{\frac{8}{3\pi}}\left(M^{1/2}\epsilon^2 \delta E \
 \right.\nonumber\\&-&\left.\frac{2}{3}~M^{3/2}\epsilon^2-\frac{1}{4}~M^{-1/2}\delta E^2\right)\, ,
\end{eqnarray}
which is clearly negative due to the dominant second and third terms. It is easy to see that the
discriminant of the above quadratic is negative and hence $\Delta J <0$, signalling no overspinning of
black hole. A five dimensional black hole with single rotation in contrast to its double rotation
counterpart cannot be overspun by linear order accretion, and thereby it obeys CCC. As stated earlier,
we have verified by following the non-linear analysis in Ref. \cite{An18} that as expected the result
continues to hold good for non-linear regime as well. 

\section{Conclusion}\label{Sec:conclusion}

We thus have that a five dimensional rotating black hole with only one rotation cannot be overspun to
create a naked singularity even for a linear order test particle accretion while the opposite is true
for the case of two rotations. For non-linear accretion, black hole can however never be overspun
whether it has one or two rotations. Further it is well known that for a black hole with one rotation
has no extremal limit in dimension $\geq 6$, and hence it could have arbitrary value of rotation
parameter without risking overspinning and destruction of horizon.

Thus we can make a general pronouncement that a higher dimensional rotating black hole with dimension
$> 4$ with only single rotation always obeys the CCC in the weak form as it could not be overspun to destroy the horizon. 

{Let us note a black hole has one rotation in four dimension while it can have two rotations in five dimension. 
For linear accretion it appears that overspinning is only possible when black hole has maximum number of possible rotations; i.e. 
one for four dimension and two for five dimension. That is, a four dimensional rotating black hole and five dimensional black hole with two rotations 
could be overspun to destroy the horizon, thereby violating CCC in the weak form. On the other hand five dimensional black hole with 
single rotation can never be overspun. And black hole with single rotation in dimension greater than five cannot anyway be overspun as 
there exists no extremal limit. }

{The main result of this work could be couched as: \it {All higher dimensional ($>4$) rotating black holes with only one single rotation can never be 
overspun under test particle linear accretion and hence would always obey CCC in the weak form.} }

\section*{Acknowledgments}
BA acknowledges the Faculty of Philosophy and Science, Silesian
University in Opava, Czech Republic, Inter-University Centre for
Astronomy and Astrophysics, Pune, India, and Goethe University,
Frankfurt am Main, Germany, for warm hospitality. This research is supported in part by Projects No. VA-FA-F-2-008 and No. MRB-AN-2019-29 of the Uzbekistan Ministry for Innovation Development, by the Abdus Salam International Centre for Theoretical Physics through Grant No. OEA-NT-01 and by an Erasmus + Exchange Grant between Silesian University in Opava and National University of Uzbekistan.

\bibliographystyle{apsrev4-1}  
\bibliography{gravreferences}

 \end{document}